\documentclass[12pt]{article}
\usepackage{amssymb}
\usepackage[dvips]{epsfig}
\newcommand{\be}{\begin{equation}}
\newcommand{\ee}{\end{equation}}
\def\bea{\begin{eqnarray}}
\def\eea{\end{eqnarray}}
\newcommand{\bn}{\begin{eqnarray}}
\newcommand{\en}{\end{eqnarray}}
\newcommand{\nn}{\nonumber}
\newcommand{\no}{\noindent}

\newcommand{\bu}{\bullet}
\newcommand{\p}{\partial}
\def \oomega {\overline \omega }
\def \oOmega {\overline \Omega }
\def \ou {\overline U }
\def \oK {\overline K }
\def \bu {\mathbb{U} }
\def \obu {\overline \mathbb{U} }
\def \oc {\overline C }
\def \ola {\overline \lambda }
\def \ooe {\overline e }
\def \of {\overline f }
\def \og {\overline g }
\def \oG {\overline G }
\def \oy {\overline y }
\def \ty {\tilde{y} }

\def \oor {R }
\def \os { S }
\def \on { N }
\def \om { M }
\def \oa {A }
\def \ob { B }
\def \te {\tilde{e} }
\def \tu {\tilde{U} }
\def \ooy {\overline Y}
\def \oxi {\overline \xi }

\def \Lag {\mathcal{L}}
\newcommand{\beq}{\begin{eqnarray}}
\newcommand{\eeq}{\end{eqnarray}}

\begin{document}

\title{\textbf{Self-dual models in $D=2+1$ from dimensional reduction}}
\author{D. Dalmazi$^{1}$\footnote{denis.dalmazi@unesp.br} \\
\textit{{UNESP - Campus de Guaratinguet\'a - Departamento 
de Fisica}}\\
\textit{{CEP 12516-410 - Guaratinguet\'a - SP - Brazil.} }\\}
\date{\today}
\maketitle

\begin{abstract}

Here we perform a Kaluza-Klein dimensional reduction of  Vasiliev's first-order description of massless spin-s particles from $D=3+1$ to $D=2+1$ and derive first-order self-dual models describing particles with helicities $\pm s$ for the cases $s=1,2,3$.  In the first two cases we recover known (parity singlets) self-dual models. In the spin-3 case we derive a new first order self-dual model with a local Weyl symmetry which lifts the traceless restriction on the rank-3 tensor. A gauge fixed version of this model corresponds to a known spin-3 self-dual model.  We conjecture that our procedure can be generalized to  arbitrary integer spins.

\end{abstract}

\newpage

\section{Introduction}

Elementary massless particles of spin-s in $D=3+1$ have well defined helicities $\pm s$ however, it is not possible to write down local Lagrangians for helicity eigenstates, they can not be locally separated. The situation is different in $D=2+1$ where massive spin-s particles with helicity $+s$ or $-s$ can be described by local Lagrangians (parity singlets), also called self dual models. In the present work the symbol $SD_j^{(s)}$ stands for a self-dual model of helicity $s$ and of $j$-th order in derivatives. Those models are irreducible representations of the Poincare group in $D=2+1$. The Maxwell-Chern-Simons theory and the linearized topologically massive gravity \cite{djt} correspond respectively to $SD_2^{(\pm 1)}$ and $SD_3^{(\pm 2)}$. 

Remarkably, self dual models of opposite helicities $+s$ and $-s$ can be ``soldered'' into a consistent (ghost free) parity doublet as in the spin-1 \cite{bk} and spin-2 \cite{iw,dmsolder1} cases and more recently for spin-3/2 \cite{mls} and spin-3 \cite{dsmb}. In particular, the soldering of second order spin-2 self dual models, as defined in \cite{desermc}, gives rise to the well known Fierz-Pauli \cite{fp} massive spin-2 theory while the soldering of linearized topologically massive gravities leads to the linearized new massive gravity (NMG) of \cite{bht}. The fine tuned coefficients necessary in \cite{fp,bht} in order to have a ghost free theory are automatically produced by the soldering procedure. 

In the case of higher spin models ($s>2$) the soldering procedure  may 
furnish interesting hints about the much less known higher spin geometry. In the spin-3 case the soldering of $SD_6^{(\pm 3)}$ (or $SD_5^{(\pm 3)}$) has produced a sixth order parity doublet model which seems to be a natural spin-3 generalization of linearized NMG. It reinforces the naturalness  of the restricted (traceless) symmetry $\delta h_{\mu\nu\rho}=\p_{(\mu}\oxi_{\nu\rho)}$ as opposed to  the non restricted one, see \cite{dsmb}. We believe that similar solderings of $SD_{2s}^{(\pm s)}$ and $SD_{2s-1}^{(\pm s)}$ into parity doublets for $s\ge 4$ can be carried out and may inspire us to develop a better understanding of higher spin geometry. An important technical point is the high number of derivatives that makes the proof of unitarity cumbersome as the spin increases. Moreover the connection with $D=3+1$ massive models is not clear since those higher derivative models are unitary only  in $D=2+1$ just like NMG.

On the other hand, if we stick to lower order self dual models other technical challenges show up. Indeed, it took  us some time until we were able to overcome the spin-2 barrier for the soldering procedure. The main obstacle is the typical presence of auxiliary fields in  higher spin theories. It is not yet clear how to include the auxiliary fields in the soldering procedure. Fortunately, thanks to the trading of auxiliary fields into higher derivatives, the higher order  self dual models used in \cite{dsmb}  do not have auxiliary fields. We still do not know how to solder $SD_j^{(\pm 3)}$ with $j=1,2,3,4$.

We believe that the previous problem is related to another one. Namely, 
for each spin $s=1,3/2,2,3$ there are $2s$ equivalent self-dual models $SD_j^{(\pm s)}$ with $ j=1,2, \dots,2s $. In the cases $s=1,3/2,2$ it is possible to go from $SD_{j}^{(s)}$ to $SD_{j+1}^{(s)}$ from bottom to top  via a Noether gauge embedding procedure (NGE). However, at $s=3$ although we can go from $SD_{j}^{(3)}$  to $SD_{j+1}^{(3)}$ for $j=1,2,3$ \cite{mdnge} and $j=5$ \cite{dss}, we can not connect $SD_{4}^{(3)}$, which contains a vector auxiliary field, with $SD_{5}^{(3)}$ which only depends on a totally symmetric rank-3 tensor. We hope that if we start from a more general first order model $SD_{1}^{(3)}$ we might be able to overcome the 4th order barrier and follow the row of models continuously until $SD_{6}^{(3)}$. 

In the present work we have been able to derive a new first order spin-3 self dual model, which generalizes the previous known model in the literature \cite{ak1}, via a Kaluza-Klein (KK) dimensional reduction of the first order Vasiliev \cite{vasiliev} description of massless spin-s models in $D=3+1$. The dimensional reduction gives rise to a pair of opposite helicities self-dual models that we can decouple via simple linear transformations (section 4), this is how  the apparent paradox of deriving helicity eigenstates (parity singlets) from four dimensional models (parity invariants) is solved.

In sections 2 and 3 we introduce our notation and basic ideas in the simpler cases of $s=1$ and $s=2$ where we reproduce known models in the literature. In section 4 we obtain the new $s=3$ self dual model and in section 5 we draw our conclusions.

%
% The infinite tower of massive modes do not mix at quadratic level so we can truncate the analysis to one massive mode. This is known to be consistent and preserve the number of degrees of freedom, see \cite{ady,rss}.
%The gauge symmetries of the massless theory are inherited in general by the massive model through the presence of Stueckelberg fields. 
%
%
%In the next sections we introduce our basic procedure in the simpler case of
%spin-1 and spin-2 particles. In section 4 we deduce a new spin-3 self-dual model via KK dimensional reduction.  In section 5 we draw our conclusions and perspectives.

\section{The spin-1 case}

The Maxwell theory can be written in a first order version with the help of an antisymmetric field $Y_{[MN]}$. In $D=3+1$-dimensions\footnote{Throughout this work we use the metric
$\eta_{\mu\nu}=(-,+,+,\cdots,+)$ and the notation:
$(\alpha\beta)=(\alpha\beta + \beta\alpha)/2$ and
$[\alpha\beta]=(\alpha\beta-\beta\alpha)/2$. } we have

\bea S_{Max}=\int{d^{4}x}\Big[\,\frac{1}{4}Y^{[MN]}Y_{[MN]}-\frac{1}{2}Y^{[MN]}(\partial_{M}e_{N}-\partial_{N}e_{M})\Big]\;.\label{max}\eea

\no Here we use capital Latin letters to denote the $3+1$-dimensional indices, ($M,N,\ldots=0,1,2,3$). In  $2+1$-dimensions we use Greek letters ($\mu,\nu,\ldots=0,1,2$). We compact the last spatial dimension $x^{D}=x^{3}\equiv y$ in a circle of radius $R=1/m$ and keep only one massive mode. This is a known \cite{ady,rss} method of obtaining massive models from massless ones.  The fields are decomposed as

\bea Y^{[MN]}(x^{\alpha},y)\rightarrow\left\{\begin{array}{l}
Y^{[\mu\nu]}(x,y)=\sqrt{\frac{m}{\pi}}\,y^{[\mu\nu]}(x)\cos{my}\\
Y^{[\mu{D}]}(x,y)=\sqrt{\frac{m}{\pi}}\,C^{\mu}(x)\sin{my}\\
e^{\mu}(x,y)=\sqrt{\frac{m}{\pi}}\,e^{\mu}(x)\cos{my}\\
e^{D}(x,y)=\sqrt{\frac{m}{\pi}}\,\phi(x)\sin{my}
\end{array}\right.
\eea

\no Integrating over $y$ in (\ref{max}) ranging from $0$ to $2\pi/m$ we obtain an action in $D=2+1$ whose Lagrangian is a first order version of the Maxwell-Proca theory, 

\be \Lag_{MP} = \frac 14 y^{[\mu\nu]}y_{[\mu\nu]} - \frac 12 y^{[\mu\nu]}(\p_{\mu}g_{\nu} - \p_{\nu}g_{\mu}) + \frac 12 C^{\mu}C_{\mu} + m\, C^{\mu}g_{\mu} \quad . \label{lmp}\ee

\no where we have introduced the $U(1)$ invariant vector field $g_{\mu} = e_{\mu} + \p_{\mu}\phi/m $.  The gauge symmetry $(\delta e_{\mu}, \delta \phi)=(\p_{\mu}\Lambda,-m\Lambda)$ is inherited from $\delta e_M= \p_M \Lambda$ in (\ref{max}). Integrating over $C_{\mu}$ and introducing another gauge invariant vector field,
 without loss of generality, 
 
 \be y_{[\mu\nu]}\equiv m\, \epsilon_{\mu\nu\rho}f^{\rho} \quad , \label{f1} \ee
  
\no after using $\epsilon$-identities\footnote{\bea \epsilon_{\mu\nu\lambda}\epsilon_{\alpha\beta\gamma} &=& -\eta_{\mu\alpha}\eta_{\nu\beta}\eta_{\lambda\gamma}+\eta_{\mu\beta}\eta_{\nu\alpha}\eta_{\lambda\gamma}- \eta_{\mu\beta}\eta_{\nu\gamma}\eta_{\lambda\alpha}+\eta_{\mu\gamma}\eta_{\nu\beta}\eta_{\lambda\alpha}- \eta_{\mu\gamma}\eta_{\nu\alpha}\eta_{\lambda\beta}+ \eta_{\mu\alpha}\eta_{\nu\gamma}\eta_{\lambda\beta} \nn\\ \epsilon^{\mu\nu\alpha}\epsilon_{\gamma\rho\alpha} &=& \delta^{\mu}_{\rho} \delta^{\nu}_{\gamma} - \delta^{\mu}_{\gamma}\delta^{\nu}_{\rho} \quad ; \quad \epsilon^{\mu\nu\alpha}\epsilon_{\gamma\nu\alpha} = -2\, \delta^{\mu}_{\gamma} \quad ; \quad \epsilon^{\mu\nu\alpha}\epsilon_{\mu\nu\alpha} =  -6 \nn\eea } we have, 

\be \Lag_{MP} = m \, f^{\rho}E_{\rho\mu}\,g^{\mu} - \frac{m^2}2 f_{\rho}^2 -
\frac{m^2}2 g_{\rho}^2 \quad . \label{lmp1b} \ee

\no Where $E_{\mu\nu} = \epsilon_{\mu\nu\rho}\p^{\rho}$. Both vector fields $(f_{\rho},g_{\rho})$ are gauge invariants. On one hand, if we integrate over $f_{\rho}$ we obtain a Stueckelberg version of the Maxwell-Proca model in terms of $g_{\mu}$.  On the other hand, after a simple rotation in the field space we can rewrite (\ref{lmp1b}),

\be \Lag_{MP} = \Lag^{(1)}_{SD1}(m,e_{\mu}^{+})+ \Lag^{(1)}_{SD1}(-m,e_{\mu}^{-})\quad . \label{lmp1} \ee

\no where\footnote{Those rotations play a similar role to the earlier canonical transformations  carried out in \cite{bk2} where the authors have started from the usual second order Maxwell-Proca theory (if $m_+=m_-=m$, in their notation).} $e^{\pm}_{\mu}=(f_{\mu} \pm g_{\mu})/\sqrt{2}$ and $\Lag^{(1)}_{SD1}$ is the spin-1 self-dual model suggested long ago in \cite{tpn},

\be \Lag^{(1)}_{SD1}(m,e) = \frac m2 e^{\mu}E_{\mu\nu}\, e^{\nu} - \frac{m^2}2 e_{\mu}^2 \quad . \label{lsd1} \ee

\no which describes massive particles of helicity $\vert m \vert/m$. Notice that the $U(1)$ symmetry in (\ref{lmp1}) can be gauged away at action level via the unitary gauge $\phi=0$ which uniquely determines the gauge parameter, thus satisfying the ``completeness'' criteria of \cite{moto}. Usually, gauge conditions can not be imposed at action level due to the loss of equations of motion. However, the ``completeness'' criteria of \cite{moto} guarantees that no relevant equation is lost. So we have been able to rewrite (\ref{lmp}) in terms of a pair of decoupled vector fields of helicities $+1$ and $-1$.

For a closer comparison between the Maxwell-Proca theory and the doublet model (\ref{lmp1}), one may check that if we add to (\ref{lmp1b}) source terms $J_f^{\mu}f_{\mu} + J_g^{\mu}g_{\mu}$ with arbitrary currents and integrate over $f_{\rho}$, we derive the Maxwell-Proca model in terms of $g_{\rho}$ with linear and quadratic source terms. Comparing  with the corresponding source terms  that would appear in (\ref{lmp1}) we are led to the map $e_{\mu}^{\pm} \leftrightarrow \left(\pm \, g_{\mu}  + E_{\mu\nu}g^{\nu}/m\right)/\sqrt{2} $. The map holds true inside correlation functions up to contact terms due to the quadratic terms in the sources. Since equations of motion of a quadratic theory like Maxwell-Proca hold true also at quantum level  up to contact terms\footnote{See the third footnote in \cite{spectator}.}, the Klein-Gordon equation $(\Box -m^2)g_{\rho}=0$  and the transverse condition $\p^{\rho}g_{\rho}=0$ are also valid at quantum level under the same circumstances. Thus,  the map is equivalent to $e_{\mu}^{\pm} \leftrightarrow \pm \sqrt{2}(P^{\pm})_{\mu\nu}g^{\nu}$ where $P^{\pm}_{\mu\nu}=\left(\theta_{\mu\nu} \pm E_{\mu\nu}/\sqrt{\Box}\right)/2$ are the 
 projection operators into $\pm 1$ helicity eigenstates in $D=2+1$, with $\Box \theta_{\mu\nu} \equiv \Box \eta_{\mu\nu} -\p_{\mu}\p_{\nu}$ and noticing $E_{\mu\nu}E^{\nu}_{\quad\!\!\!\! \beta} = \Box\theta_{\mu\beta}$.

\section{The spin-2 case}

Massless spin-2 particles are commonly described by the linearized Einstein-Hilbert (EH) theory in terms of a symmetric rank-2 tensor field. This theory can be written in a first order version using a non-symmetric rank-2 tensor, $e_{MN}$, and a mixed symmetry rank-3 tensor $Y^{[AB]M}$ as in the spin-2 case of the Vasiliev's  formulation of massless spin-s particles \cite{vasiliev}, using the notation of \cite{bch} we have

\bea S_{s=2}&=&\frac{1}{2}\int{d^{4}x}\Big[Y^{[AB]M}Y_{[AM]B}-\frac{Y^{A}Y_{A}}{2}-Y^{[AB]M}(\partial_{A}e_{BM}-\partial_{B}e_{AM})\Big]\;,\label{s=2}\eea

\no where $Y^{B}=\eta_{MA}Y^{[AB]M}$. The action $S_{s=2}$ is invariant under the following gauge transformations:

\bea \delta e_{AB} &=& \p_A \xi_B + \oomega_{[AB]} \, , \label{deltae1} \\ \delta Y_{[BA]M} &=& \p_M \oomega_{[AB]} + \eta_{AM} \p^C \oomega_{[BC]} - \eta_{BM}\p^C\oomega_{[AC]} \, \label{deltaw1} \eea

The Gaussian integrals over the auxiliary fields $Y_{[MA]B}$ lead to the linearized Einstein-Hilbert theory in terms of $e_{(MN)}$. The KK dimensional reduction of $S_{s=2}$ has been carried out in \cite{kmu} but for the sake of comparison with the more involved spin-3 case we reproduce here some formulae in a convenient notation.  Compactfying the spatial dimension $y=x^{D}$ in a circle as in the spin-1 case, the fields and the gauge parameters are redefined according to:

\bea Y^{[AB]M}(x,y)&\rightarrow&\left\{\begin{array}{l}
Y^{[\alpha\beta]\mu}=\sqrt{\frac{m}{\pi}}\,y^{[\alpha\beta]\mu}(x)\cos{my}\\
Y^{[\alpha\beta]D}=\sqrt{\frac{m}{\pi}}\,y^{[\alpha\beta]}(x)\sin{my}\\
Y^{[\alpha{D}]\mu}=\sqrt{\frac{m}{\pi}}\,C^{\mu\alpha}(x)\sin{my}\\
Y^{[\alpha D]D}=\sqrt{\frac{m}{\pi}}\,C^{\alpha}(x)\cos{my}
\end{array}\right.\;,\label{Y2}\\
\nn\\
e^{MN}(x,y)&\rightarrow&\left\{\begin{array}{l}
e^{\mu\nu}(x,y)=\sqrt{\frac{m}{\pi}}e^{\mu\nu}(x)\cos{my}\\
e^{\mu D}(x,y)=\sqrt{\frac{m}{\pi}}U^{\mu}(x)\sin{my}\\
e^{D\nu}(x,y)=\sqrt{\frac{m}{\pi}}S^{\nu}(x)\sin{my}\\
e^{DD}(x,y)=\sqrt{\frac{m}{\pi}}\phi(x)\cos{my}\\
\end{array}\right.\;,\label{e2}\\
\nn\\
\xi^{M}(x,y)&\rightarrow&\left\{\begin{array}{l}
\xi^{\mu}(x,y)=\sqrt{\frac{m}{\pi}}\epsilon^{\mu}(x)\cos{my}\\
\xi^{D}(x,y)=\sqrt{\frac{m}{\pi}}\epsilon(x)\sin{my}\\
\end{array}\right.\;,\label{xi2}\\
\nn\\
\oomega^{[MN]}(x,y)&\rightarrow&\left\{\begin{array}{l}
\oomega^{[\mu\nu]}(x,y)=\sqrt{\frac{m}{\pi}}\oomega^{[\mu\nu]}(x)\cos{my}\\
\oomega^{[\mu{D}]}(x,y)=\sqrt{\frac{m}{\pi}}\oomega^{\mu}(x)\sin{my}\\
\end{array}\right.\;,\label{omega2}\eea

\no where $C_{\mu\nu}$ is an arbitrary rank-2 tensor without symmetry. Once again we introduce a dual non symmetric rank-2 field $f_{\rho\mu}$ via the Levi-Civita tensor, without loss of generality,

\be y_{[\alpha\beta]\mu} \equiv m \, \epsilon_{\alpha\beta}^{\quad\rho}f_{\rho\mu} \quad . \label{f2} \ee

 After integrating over the cyclic coordinate $x_D=y$ we obtain a 3D massive action whose Lagrangian is given by

\bea \Lag_m^{(2)} &=& -m\, f_{\rho\gamma}E^{\rho\beta}e_{\beta}^{\,\,\gamma} + \frac{m^2}2 (f_{\alpha\beta}f^{\beta\alpha} - f^2) + y^{[\mu\nu ]}\p_{\mu}\ou_{\nu} \nn\\ &-&  \frac 12 C^{\mu\nu}C_{\nu\mu} + \frac{C^2}4 + C^{\mu\nu}\left( m\, e_{\nu\mu}+\p_{\nu}S_{\mu} - y_{[\nu\mu]} \right) \nn \\ &-&\frac 14 C_{\mu}^2 + m\, C_{\mu} \left( \ou^{\mu}-\frac{\epsilon^{\mu\alpha\beta}f_{\alpha\beta}}2\right) \label{lm2c}\eea

\no Where $\ou_{\mu} = U_{\mu} - \p_{\mu}\phi/m $. Integrating over the C-fields we are able to write down the Lagrangian in a simpler form:

\be \Lag_m^{(2)} = -m\, f^{*\rho\gamma}E_{\rho}^{\,\,\beta}e^*_{\beta\gamma}+ \frac{m^2}2 \left\lbrack f^*_{\alpha\beta}f^{*\beta\alpha} - (f^*)^2\right\rbrack + \frac{m^2}2 \left\lbrack e^*_{\alpha\beta}e^{*\beta\alpha} - (e^*)^2\right\rbrack \label{lm2*} \ee 

\no with

\bea f^*_{\mu\nu} &=& f_{\mu\nu}  - \frac 1m\p_{\mu}K_{\rho} + \epsilon_{\mu\nu\rho}\ou^{\rho}   \label{fstar}  \\
e^*_{\mu\nu} &=& e_{\mu\nu}  + \frac 1m\p_{\mu}S_{\rho} - \frac 1m\p_{\mu}\ou_{\rho}+\epsilon_{\mu\nu\rho}K^{\rho} \, . \label{estar} \eea

\no where we have introduced the dual $K_{\rho}$ field via the invertible map:

\be y_{[\mu\nu]} = m \, \epsilon_{\mu\nu\rho}K^{\rho} \quad . \label{K2}\ee

Finally, after a simple rotation we can disentangle the helicity $\pm 2$ eigenstates from (\ref{lm2*}) such that

\be \Lag_m^{(2)} = \Lag^{(2)}_{SD1}(m,e^{+})+ \Lag^{(2)}_{SD1}(-m,e^{-})\quad , \label{lmp2} \ee

\no  where $e_{\mu\nu}^{(\pm)} = (e^{*}_{\mu\nu} \pm f^*_{\mu\nu})/\sqrt{2}$. We obtain the spin-2 self-dual model of Aragone and Khoudeir \cite{ak} with  helicity $2\,\vert m \vert/m$, i.e., 

\be \Lag^{(2)}_{SD1}(m,e) = \frac m2\, e^{\rho\gamma}E_{\rho}^{\,\,\beta}e_{\beta\gamma}+ \frac{m^2}2 \left\lbrack e_{\alpha\beta}e^{\beta\alpha} - e^2\right\rbrack \label{lsd2} \ee

\no As a double check of our final Lagrangian, it is easy to show that  (\ref{lm2*}) and consequently (\ref{lmp2}), is invariant under the gauge symmetries associated with the parameters (\ref{xi2}) and (\ref{omega2}), i.e., 

\bea \delta f_{\alpha\beta} &=& \epsilon_{\alpha\beta\rho}\oomega^{\rho} - \p_{\alpha}\omega_{\beta}/m \quad ; \quad \delta K_{\rho} = - \omega_{\rho}  \label{deltafk2} \\
\delta e_{\alpha\beta} &=& \epsilon_{\alpha\beta\rho}\omega^{\rho} + \p_{\alpha}\epsilon_{\beta} \quad ; \quad \delta S_{\rho} = -m\, \epsilon_{\rho} - \oomega_{\rho} \label{deltaes2} \\
\delta U_{\rho} &=& \oomega_{\rho} + \p_{\rho} \epsilon  \quad ; \quad \delta \phi = m\, \epsilon \quad , \label{deltauphi} \eea

\no where $\omega_{\rho} \equiv -\epsilon_{\rho\mu\nu}\oomega^{\mu\nu}/2$. 
The symmetries follow from the gauge invariance of the composite fields $f_{\mu\nu}^*$ and $e_{\mu\nu}^*$. This will not be true in the spin-3 case as will see in the next section. We can turn the composite fields into elementary ones after fixing  the unitary gauge at action level:

\be (K_{\rho},U_{\rho},S_{\rho},\phi)=(0,0,0,0)  \quad . \label{ug2} \ee

\no Once $f_{\mu\nu}^*$ and $e_{\mu\nu}^*$ are considered elementary fields,  no symmetry is left in (\ref{lm2*}). The fact that we are allowed to fix (\ref{ug2}) at action level is grounded on the ``completeness'' criteria, see \cite{moto}. Namely, the 10 gauge conditions (\ref{ug2}) uniquely (completely)  determine the same number of gauge parameters: $(\omega_{\rho},\oomega_{\rho},\epsilon_{\rho},\epsilon)$.

\section{The spin-3 case}

The Vasiliev's model \cite{vasiliev} for a massless spin-3 particle, in the notation of \cite{bch}, is given by

\bea \Lag_{s=3} &=& \frac{\ooy^{[MN](\oor\os)}\ooy_{[MN](\oor\os)}}3-\frac{4\, \ooy^{[MN](\oor\os)}\ooy_{[RN](\om\os)}}3 + \frac{4\, \ooy^{[RM](\on}_{\quad\,\,\,\,\quad\oor)} \ooy_{[SN](\om}^{\quad\,\,\,\,\quad\os)}}9 \nn\\ &+& \ooy^{[MN](\oor\os)}\p_M \ooe_{N(\oor\os)}   \label{ls3} \eea

\no The bars remind us of the traceless conditions:

\be \eta^{RS}\ooy_{[MN](\oor\os)}=0=\eta^{RS}\ooe_{N(\oor\os)} \label{tr1} \ee

\no The 4D action corresponding to (\ref{ls3}) is invariant under the gauge symmetries:

\bea \delta \ooe_{N(\oor\os)} &=& \p_N \oxi_{(\oor\os)} + \oomega_{\on(\oor\os)}
\quad , \label{deltae3} \\
(-4)\, \delta \ooy_{[MN](\oor\os)} &=& \p_{R} \oomega_{[\om\on]\os}+ \p_{S} \oomega_{[\om\on]\oor}+ \eta_{MR} \p^A \oomega_{[\on\oa]\os} + \eta_{MS} \p^A \oomega_{[\on\oa]\oor}\nn\\ &-& \eta_{NR} \p^A \oomega_{[\om\oa]\os} - \eta_{NS} \p^A \oomega_{[\om\oa]\oor} , \label{deltay3a} \eea
\no where $\oomega_{[\om\on]\os} = (\oomega_{\om(\on\os)}-\oomega_{\on(\om\os)})/2$ and we have:

\bea \eta^{RS}\oxi_{(\oor\os)} &=& 0 \quad , \label{trxi}\\
 \eta^{RS}\oomega_{M(\oor\os)}&=& 0 = \eta^{MR}\oomega_{M(\oor\os)} \label{tromega} \eea

\no and also

\be  \oomega_{M(\oor\os)} + \oomega_{R(\os \om)} + \oomega_{S(\om\oor)} =0 \quad . \label{omegac} \ee

The dimensional reduction from $3+1$ to $2+1$ is performed as before with the notation:

\bea \ooy_{[MN](\oa\ob)}(x,y)&\rightarrow&\left\{\begin{array}{l}
Y^{[\mu\nu](\alpha\beta)}(x,y)=\sqrt{\frac{m}{\pi}}\,y^{[\mu\nu](\alpha\beta)}(x)\cos{my}\\
Y^{[\mu\nu](\alpha D)}(x,y)=\sqrt{\frac{m}{\pi}}\,y^{\alpha[\mu\nu]}(x)\sin{my}\\
Y^{[\mu D](\alpha \beta)}(x,y)=\sqrt{\frac{m}{\pi}}\,C^{\mu(\alpha\beta)}(x)\sin{my}\\
Y^{[\mu D](\alpha D)}(x,y)=\sqrt{\frac{m}{\pi}}
\,C^{\mu\alpha}(x)\cos{my}\\
\end{array}\right.\;,\label{y3}\\
\nn\\
\ooe^{M(\oa\ob)}(x,y)&\rightarrow&\left\{\begin{array}{l}
e^{\mu(\alpha\beta)}(x,y)=\sqrt{\frac{m}{\pi}}
e^{\mu(\alpha\beta)}(x)\cos{my}\\
e^{\mu(\alpha D)}(x,y)=\sqrt{\frac{m}{\pi}}
U^{\mu\alpha}(x)\sin{my}\\
e^{D(\alpha\beta)}(x,y)=\sqrt{\frac{m}{\pi}}
S^{(\alpha\beta)}(x)\sin{my}\\
e^{D(D\alpha)}(x,y)=\sqrt{\frac{m}{\pi}}
\phi^{\alpha}(x)\cos{my}\\
\end{array}\right.\;,\label{e3}\\
\nn\\
\oxi^{(\oa\ob)}(x,y)&\rightarrow&\left\{\begin{array}{l}
\xi^{(\alpha\beta)}(x,y)=\sqrt{\frac{m}{\pi}}\epsilon^{(\alpha\beta)}(x)\cos{my}\\
\xi^{(\alpha D)}(x,y)=\sqrt{\frac{m}{\pi}}\epsilon^{\alpha}(x)\sin{my}\\
\end{array}\right.\;,\label{xi3}\\
\nn\\
\oomega_{M(\oa\ob)}(x,y)&\rightarrow&\left\{\begin{array}{l}
\omega^{\mu(\alpha\beta)}(x,y)=\sqrt{\frac{m}{\pi}}\omega^{\mu(\alpha\beta)}(x)\cos{my}\\
\omega^{\alpha(D\beta)}(x,y)=\sqrt{\frac{m}{\pi}}\oomega^{\alpha\beta}(x)\sin{my}\\
\end{array}\right.\;,\label{omega3}\eea
where $C_{\mu\nu}$, $U_{\mu\nu}$ and $\oomega_{\mu\nu}$  are non symmetric rank-2 tensors, the last one is traceless:

\be \eta^{\alpha\beta}\oomega_{\alpha\beta} =0 \quad . \label{tr3} \ee

Now several words are in order before we proceed. The reader may be missing the parameters $ \omega_{D(DD)},\omega_{D(D\alpha)},\omega_{\alpha(DD)},\xi_{(DD)}$. The first one vanishes as one can see by fixing  $(M,R,S)=(D,D,D)$ in (\ref{omegac}). The other ones are not independent quantities, due to the traceless conditions and (\ref{omegac})  we have $(\omega_{D(D\alpha)},\omega_{\alpha(DD)},\xi_{(DD)})$ =$(-\oomega_{\alpha}, -\omega_{\alpha}, - \eta^{\mu\nu}\epsilon_{(\mu\nu)})$ where $
\oomega_{\alpha}\equiv \eta^{\mu\nu}\omega_{\mu(\nu\alpha)}$ and $\omega_{\alpha}\equiv\eta^{\mu\nu}\omega_{\alpha(\mu\nu)}$. From 
(\ref{omegac}) we also have the traceless condition (\ref{tr3}),
$ \oomega_{\alpha} = - \omega_{\alpha}/2 $ and

\be \omega_{\mu(\nu\alpha)} + \omega_{\nu(\alpha\mu)} + \omega_{\alpha(\mu\nu)} =0 \quad . \label{omegac2} \ee

\no For the accounting of the number of independent gauge parameters we notice that the 10 constraints (\ref{omegac2}) allow us to write the 18 gauge parameters $\omega_{\mu(\nu\alpha)}$ in terms of a non symmetric traceless rank-2 tensor $\oOmega_{\mu\nu}$ with 8 independent degrees of freedom. Indeed, using $\epsilon$-identities one can can show: $\omega_{\alpha(\beta\gamma)}= \left\lbrack\epsilon_{\alpha\beta}^{\quad\rho} \left(\epsilon_{\rho}^{\quad\nu\lambda}\omega_{\lambda(\nu\gamma)}\right) + \epsilon_{\alpha\gamma}^{\quad\rho} \left(\epsilon_{\rho}^{\quad\nu\lambda}\omega_{\lambda(\nu\beta)}\right)\right\rbrack/3 $. This suggests that we can always rewrite $\omega_{\alpha(\beta\gamma)}$ in terms of a traceless rank-2 tensor. Indeed, 

\be \omega_{\alpha(\beta\gamma)} = \epsilon_{\alpha\beta}^{\quad\rho}\oOmega_{\rho\gamma} + \epsilon_{\alpha\gamma}^{\quad\rho}\oOmega_{\rho\beta} \quad . \label{omegabar} \ee

\no solves (\ref{omegac2}) if $\eta^{\rho\beta}\oOmega_{\rho\beta}=0$. In particular, $\omega_{\mu}=-2\epsilon_{\mu\alpha\beta}\oOmega^{\alpha\beta}$ only depends on the 3 antisymmetric components $\oOmega^{[\alpha\beta]}$.

After integrating over the cyclic coordinate $y$ we have a massive spin-3 theory in $D=2+1$ dimensions whose Lagrangian is given by

\bea \Lag_{m}^{(3)} &=& y^{[\mu\nu](\alpha\beta)}\p_{\mu}\te_{\nu(\alpha\beta)} + \frac 13 \left\lbrack y_{[\mu\nu](\alpha\beta)}^2-4 y_{[\mu\nu](\alpha\beta)}y^{[\mu\alpha](\nu\beta)}+y_{[\mu\nu]}^2\right\rbrack +\frac 49 y_{\alpha\beta}y^{\beta\alpha} \nn\\
&+& \frac 23 y_{\mu[\alpha\beta]}^2 - \frac 43 y_{\mu[\alpha\beta]}y^{\alpha[\mu\beta]}+ 2 \, y^{\mu[\alpha\beta]} \p_{\alpha}\tu_{\beta\mu} +\Lag_{C_{\mu\nu}} + \Lag_{C_{\mu(\nu\rho)}} \label{ls3m} \eea

\no where

\bea \te_{\mu(\alpha\beta)} = e_{\mu(\alpha\beta)} + \p_{\mu}S_{(\alpha\beta)}/m + \eta_{\alpha\beta}(e_{\mu} + \p_{\mu}S/m) \quad ; \quad e_{\mu} = \eta^{\alpha\beta}e_{\mu(\alpha\beta)} \,\, ; \,\, S = \eta^{\alpha\beta} S_{(\alpha\beta)}  , \label{etil} \\
y_{[\mu\nu]}= \eta^{\alpha\beta}y_{[\mu\nu](\alpha\beta)} \quad ; \quad y_{\mu\beta}= \eta^{\nu\alpha} y_{[\mu\nu](\alpha\beta)} \quad ; \quad \tu_{\mu\nu} = U_{\mu\nu} - \p_{\mu}\phi_{\nu}/m \, , \label{util} \eea

\bea \Lag_{C_{\mu(\nu\rho)}} &=& \frac 23 C_{\mu(\nu\rho)}^2 - \frac 43 C_{\mu(\nu\rho)}C^{\nu(\mu\rho)}- \frac 23 C_{\mu}^2 + \frac 89 C_{\mu}\oc^{\mu} + m\, C^{\mu(\nu\rho)}g_{\mu(\nu\rho)} , \label{c3} \\
\Lag_{C_{\mu\nu}} &=& - \frac 89 C_{\mu\nu}C^{\nu\mu} + \frac 49 C^2 - C_{\mu\nu}T^{\mu\nu} \quad . \label{c2} \eea 

\no with 

\be g_{\mu(\alpha\beta)} = \te_{\mu(\alpha\beta)} + \frac{4}{9\, m}\left(\eta_{\mu\alpha} y_{\beta}+\eta_{\mu\beta}y_{\alpha} - 3 y_{\beta[\mu\alpha]} - 3y_{\alpha[\mu\beta]} \right) \, \label{g} \ee
\be T_{\mu\nu} = m\, \tu_{\mu\nu} + \frac 49 y_{\nu\mu} + \frac 43 y_{[\nu\mu]} \quad ; \quad y_{\mu} = \eta^{\alpha\beta}y_{\alpha[\beta\mu]} \quad ; \quad C_{\mu} = \eta^{\alpha\beta}C_{\mu(\alpha\beta)} \quad ; \quad 
\oc_{\beta} = \eta^{\mu\alpha}C_{\mu(\alpha\beta)} , \label{t} \ee

\no Performing the Gaussian integrals over $C_{\mu\nu}$ and $C_{\mu(\nu\rho)}$ in (\ref{ls3m}) amounts to the replacement:

\bea \Lag_{C_{\mu\nu}} + \Lag_{C_{\mu(\nu\rho)}} \to \frac 98 \left(T_{\mu\nu}T^{\nu\mu} - T^2 \right) + \frac{3\, m^2}8 \left\lbrack g^{\mu(\alpha\nu)}g_{\alpha(\mu\nu)}-\og_{\mu}\og^{\mu} + \frac{g_{\mu}g^{\mu}}4 \right\rbrack , \label{replace} \eea

\no where $g_{\mu} = \eta^{\alpha\beta}g_{\mu(\alpha\beta)}$ and 
$\og_{\beta} = \eta^{\mu\alpha}g_{\mu(\alpha\beta)}$. The Lagrangian (\ref{ls3m}), using (\ref{replace}), is invariant under the gauge transformations:

\bea (-4)\delta y_{[\mu\nu](\alpha\beta)} &=& m\, \eta_{\mu\alpha}\left\lbrack \oomega_{[\nu\beta]} + 3 \oomega_{(\nu\beta)}\right\rbrack - 
m\, \eta_{\nu\alpha}\left\lbrack \oomega_{[\mu\beta]} + 3 \oomega_{(\mu\beta)}\right\rbrack  + (\alpha \leftrightarrow \beta ) \nn\\
&+& \eta_{\mu\alpha}\p^{\lambda}\left\lbrack \omega_{\nu(\lambda\beta)} - \omega_{\lambda(\nu\beta)}\right\rbrack - \eta_{\nu\alpha}\p^{\lambda}\left\lbrack \omega_{\mu(\lambda\beta)} - \omega_{\lambda(\mu\beta)}\right\rbrack + (\alpha \leftrightarrow \beta ) \nn\\
&+& \p_{\alpha} \left\lbrack \omega_{\mu(\nu\beta)} - \omega_{\nu(\mu\beta)} \right\rbrack + \p_{\beta} \left\lbrack \omega_{\mu(\nu\alpha)} - \omega_{\nu(\mu\alpha)} \right\rbrack \, , \label{deltay4} \eea

\bea (-4)\delta y_{\alpha[\mu\nu]} &=& m\, [\omega_{\nu(\mu\alpha)} - \omega_{\mu(\nu\alpha)}] + \frac{3m}2 (\eta_{\mu\alpha}\omega_{\nu}- \eta_{\nu\alpha}\omega_{\mu}) \nn\\ &+& 2 \, \p_{\alpha}\oomega_{[\mu\nu]} + 2\, \left(\eta_{\mu\alpha}\p^{\lambda}\oomega_{[\nu\lambda]} - \eta_{\nu\alpha}\p^{\lambda}\oomega_{[\mu\lambda]} \right)  \label{deltay3} \eea

\bea \delta\phi_{\mu} &=& m\, \epsilon_{\mu} + \omega_{\mu}/2 \quad ; \quad \delta U_{\nu\beta} = \oomega_{\nu\beta} + \p_{\nu}\epsilon_{\beta} \, , \label{deltaphiu} \\
\delta S_{(\alpha\beta)} &=& - m \epsilon_{(\alpha\beta)}- 2 \oomega_{(\alpha\beta)} \quad ; \quad \delta e_{\mu(\alpha\beta)} = \p_{\mu} \epsilon_{(\alpha\beta)} + \omega_{\mu(\alpha\beta)} \, , \label{deltase} \eea 

\no where the gauge parameters must satisfy (\ref{tr3}) and (\ref{omegac2}), or (\ref{omegabar}).

Although, $g_{\mu(\alpha\beta)}$ is not fully gauge invariant, the reader can check from  (\ref{deltay3}), (\ref{deltaphiu}) and (\ref{deltase}) that $\delta g_{\mu(\alpha\beta)}$ only depends upon derivatives of $\oomega_{\mu\nu}$, the gauge parameters $\epsilon_{(\alpha\beta)}$ and $\omega_{\mu(\alpha\beta)}$ drop out. Analogously, based on (\ref{deltay4}) and (\ref{deltaphiu}) we are led to replace $y_{[\mu\nu](\alpha\beta)}$ by a new field whose gauge transformations only depend upon derivatives of $\omega_{\mu(\alpha\beta)}$ without any dependence on $\epsilon_{\mu}$ or $\oomega_{[\mu\nu]}$, namely

\bea \ty_{[\mu\nu](\alpha\beta)} &=& y_{[\mu\nu](\alpha\beta)} + \frac m4 \left\lbrack \eta_{\mu\alpha} \left( \bu_{[\nu\beta]}+ 3 \bu_{(\nu\beta)}\right) - \eta_{\nu\alpha} \left( \bu_{[\mu\beta]}+ 3 \bu_{(\mu\beta)}\right) + (\alpha \leftrightarrow \beta ) \right\rbrack \label{ty}\\ \bu_{\mu\nu} &=& U_{\mu\nu} - \frac{\p_{\mu}\phi_{\nu}}m \label{bu}  \eea

\no Similar to the spin-1 and spin-2 cases, see (\ref{f1}) and (\ref{f2}),
in order to have a more symmetric action with respect to the $\pm 3$ helicities, it is convenient to introduce, without loss of generality, the following invertible field redefinitions

%\bea \ty_{[\mu\nu](\alpha\beta)} &=& m\, \epsilon_{\mu\nu}^{\quad\,\rho} f_{\rho(\alpha\beta)} \quad , \left( \to f_{\rho(\alpha\beta)} = -\frac 1{2m} \epsilon_{\rho}^{\quad\, \mu\nu}\ty_{[\mu\nu](\alpha\beta)}\right) \label{f3} \\
%y_{\alpha[\mu\nu]} &=& m\, \epsilon_{\mu\nu}^{\quad\,\,\rho} K_{\rho\alpha} \quad , \left( \to  K_{\rho\alpha} = -\frac 1{2m} \epsilon_{\rho}^{\quad\, \mu\nu} y_{\alpha[\mu\nu]}\right)  \label{k2} \eea

\bea \ty_{[\mu\nu](\alpha\beta)} &\equiv & m\, \epsilon_{\mu\nu}^{\quad\,\rho} f_{\rho(\alpha\beta)} \quad ,  \label{f3} \\
y_{\alpha[\mu\nu]} &\equiv & m\, \epsilon_{\mu\nu}^{\quad\,\,\rho} K_{\rho\alpha} \quad ,  \label{k2} \eea

\no The non symmetric rank-2 tensor $K_{\mu\nu}$ plays a similar role as  $\bu_{\mu\nu}$. Now we can write down the parity doublet Lagrangian (\ref{ls3m}), using (\ref{tr3}), in a quite symmetric form,

\bea \Lag_{m}^{(3)} &=& - m\, f^{\rho}_{\,\,\, (\alpha\beta)} E_{\rho\mu} G^{\mu(\alpha\beta)} - m\, f^{\rho}E_{\rho\mu}G^{\mu} \nn\\
&+& \frac{2\, m^2}3 \left\lbrack f_{\mu(\alpha\beta)}f^{\alpha(\mu\beta)} - \of_{\mu}^2 + 4\, f_{\mu}^2 \right\rbrack + \frac{3\, m^2}8 \left\lbrack G_{\mu(\alpha\beta)}G^{\alpha(\mu\beta)} - \oG_{\mu}^2 + 4\, G_{\mu}^2 \right\rbrack \nn \\
&+& \frac {8\,m}9 \left( K_{[\nu\beta]} + 3 K_{(\nu\beta)}\right) \left( \p^{\nu} f^{\beta}+ \p^{\nu}\of^{\beta}- \p^{\beta} f^{\nu} - \p_{\alpha}f^{\nu(\alpha\beta)}\right)
 \label{lsm3b}\\
&+& \frac m2 \left( \bu_{[\nu\beta]} + 3 \bu_{(\nu\beta)}\right) \left( \p^{\nu} G^{\beta}+ \p^{\nu}\oG^{\beta}- \p^{\beta} G^{\nu} - \p_{\alpha}G^{\nu(\alpha\beta)}\right) \nn \\
&-& \frac 34 m^2\bu^2 - \frac 43 m^2 K^2 + \frac{2\, m}3 \left\lbrack \bu \epsilon_{\mu\nu\rho}\p^{\mu}K^{\nu\rho} + K\epsilon_{\mu\nu\rho}\p^{\mu}\bu^{\nu\rho} \right\rbrack
 \quad , \nn\eea

\no where we have made another invertible field redefinition $G_{\mu(\alpha\beta)} \equiv g_{\mu(\alpha\beta)} - \eta_{\alpha\beta}g_{\mu}/4$ in order that the $m^2 G^2$ and  $m^2 f^2 $ terms acquire the same form, see (\ref{replace}). No similar field redefinition was necessary in the previous spin-1 and spin-2 cases where the mass square terms have naturally appeared in a symmetric form.

Now we can easily decouple the $+3$ and $-3$ helicities via a trivial dilatation and  a rotation of the fields. Namely, 

\be \Lag_m^{(3)} = \Lag^{(3)}_{SD1}(m,e^{+},\lambda^{+})+ \Lag^{(3)}_{SD1}(-m,e^{-},\lambda^{-})\quad , \label{lmp4} \ee

\no where

\bea \Lag^{(3)}_{SD1}(m,e,\lambda) &=&\!\!\! - \frac m2\, e^{\rho}_{\,\,\, (\alpha\beta)} E_{\rho\mu} e^{\mu(\alpha\beta)} - \frac m2\, e^{\rho}E_{\rho\mu}e^{\mu} + \frac{m^2}2 \left\lbrack e_{\mu(\alpha\beta)}e^{\alpha(\mu\beta)} - \ooe_{\mu}^2 + 4\, e_{\mu}^2 \right\rbrack \nn\\
&+&\!\!\! m^2\lambda^2 + \frac{2\,m}3 \lambda \, \epsilon_{\mu\nu\rho}\p^{\mu}\lambda^{\nu\rho} + \frac {2\,m}3 \left( \lambda_{[\nu\beta]} + 3 \lambda_{(\nu\beta)}\right) \left\lbrack \p^{\nu} e^{\beta}+ \p^{\nu}\ooe^{\beta}- \p^{\beta} e^{\nu} - \p_{\alpha}e^{\nu(\alpha\beta)}\right\rbrack , \nn\\\label{lelambda} \eea

\no with $\lambda = \eta^{\mu\nu}\lambda_{\mu\nu}$ and 

\be e^{\pm}_{\mu(\nu\rho)} = \frac{4 \, f_{\mu(\nu\rho)} \pm 3\, G_{\mu(\nu\rho)}}{\sqrt{6}} \quad ; \quad \lambda_{\mu\nu}^{\pm} = \frac{3\, \bu_{\mu\nu} \pm 4\, K_{\mu\nu} }{\sqrt{6}} \quad , \label{6.4} \ee

\no Inspired by the first order spin-3 self-dual model of \cite{ak1}, we have further simplified (\ref{lelambda}) via

\be e_{\mu(\nu\rho)} \to e_{\mu(\nu\rho)} +\frac 2m \p_{\mu}\ola_{\nu\rho} - \frac 13 \left( \epsilon_{\mu\nu}^{\quad \beta}\lambda_{\beta\rho} +  \epsilon_{\mu\rho}^{\quad \beta}\lambda_{\beta\nu} \right) \quad , \label{6.5} \ee

\no with $\ola_{\mu\nu} = \lambda_{\mu\nu} - \eta_{\mu\nu}\lambda/3$. Consequently, the derivative couplings $\lambda \partial e$ are replaced by non derivative ones leading to our main result, a new spin-3 self dual model:

\bea \Lag^{(3)}_{SD1}(m,e,A,\phi) \!\!\! &=& \!\!\! - \frac m2\, e^{\rho}_{\,\,\, (\alpha\beta)} E_{\rho\mu} e^{\mu(\alpha\beta)} - \frac m2\, e^{\rho}E_{\rho\mu}e^{\mu} + \frac{m^2}2 \left\lbrack e_{\mu(\alpha\beta)}e^{\alpha(\mu\beta)} - \ooe_{\mu}^2 + 4\, e_{\mu}^2 \right\rbrack \nn\\
&-& \!\!\! m^2 \, (\ooe_{\mu} + 5\, e_{\mu})A^{\mu} + 3\, m^2 \, A_{\mu}^2 + m\, A^{\mu}E_{\mu\nu}A^{\nu} - \frac{4m}3 \phi \, \p^{\mu}A_{\mu} - m^2 \phi^2 \, , \nn\\ \label{6.6} \eea

\no where 

\be A_{\mu}^{\pm} \equiv  - \frac 12 \epsilon_{\mu\nu\rho}\lambda^{\nu\rho}_{\pm} \quad ; \quad \phi_{\pm} \equiv \lambda_{\pm} \label{6.7} \ee 

\no The doublet Lagrangian can be written as 

\be \Lag_m^{(3)} = \Lag^{(3)}_{SD1}(m,e^+,A_+,\phi_+) + \Lag^{(3)}_{SD1}(-m,e^-,A_-,\phi_-) \quad . \label{doublet} \ee

\no Taking into account the several field redefinitions that we have carried out so far, the composite fields in (\ref{doublet}) are given by

\bea e_{\mu(\alpha\beta)}^{\pm} &=& 4\, f_{\mu(\alpha\beta)} \pm 3 G_{\mu(\alpha\beta)} + 2\eta_{\alpha\beta} A_{\mu}^{\pm} - \eta_{\mu\beta}A_{\alpha}^{\pm} - \eta_{\mu\alpha} A_{\beta}^{\pm} \quad , \label{6.8} \\
f_{\mu(\alpha\beta)} &=& - \frac 1{2\,m} \epsilon_{\mu}^{\,\,\,\rho\nu} y_{[\mu\nu](\alpha\beta)} + \frac 2m \p_{\mu}\oK_{\alpha\beta} - \frac 14 \left\lbrack \epsilon_{\mu\alpha}^{\quad \nu}(\bu_{[\nu\beta]} + 3 \bu_{(\nu\beta)}) + (\alpha \leftrightarrow \beta ) \right\rbrack , \label{6.9} \\
G_{\mu(\alpha\beta)} &=& e_{\mu(\alpha\beta)}^{(0)} + \frac{\p_{\mu}S_{(\alpha\beta)}}m - \frac 43 (\epsilon_{\mu\alpha}^{\quad\rho}K_{\rho\beta} + \epsilon_{\mu\beta}^{\quad \rho}K_{\rho\alpha} ) \nn\\
&+& \frac 49 (\eta_{\mu\alpha}\epsilon_{\beta\lambda\rho} + \eta_{\mu\beta}\epsilon_{\alpha\lambda\rho}-2\, \eta_{\alpha\beta} \epsilon_{\mu\lambda\rho})K^{\lambda\rho} + \frac 2m \p_{\mu}\obu_{(\alpha\beta)} \label{6.10} \eea

\no where $\obu_{\mu\nu}= \bu_{\mu\nu} - \eta_{\mu\nu}\bu/3$ and 
$\oK_{\mu\nu} = K_{\mu\nu} - \eta_{\mu\nu}\,K/3$. 

 From  (\ref{deltay3}), (\ref{deltaphiu}), (\ref{k2}), (\ref{6.4}) and (\ref{6.7}) we have the gauge transformations of $A_{\mu}^{\pm}$ and $\phi^{\pm}$. Remarkably, they only depend on one vector parameter $\Lambda_{\mu}$. In the $+3$ helicity case\footnote{In formulae (\ref{deltaA})-(\ref{deltaleb}) we suppress the upper index $+$. For the general case 
$\pm 3$ we have $\Lambda_{\mu}^{\pm} = \pm \frac 32 \omega_{\mu} - \epsilon_{\mu\alpha\beta}\oomega^{[\alpha\beta]} $ and $m\to \pm m$ in 
$\delta A_{\mu}^{\pm} $ and $\delta\phi^{\pm}$} we have:

\bea \delta\phi &=& -\frac 1m \p^{\mu}\Lambda_{\mu} \quad ; \quad \delta A_{\mu} = \frac 32 \, \Lambda_{\mu} + \frac 1{2\,m} \epsilon_{\mu\alpha\beta}\p^{\alpha}\Lambda^{\beta} \, , \label{deltaA} \\ 
\Lambda_{\mu} &\equiv &  \frac 32 \omega_{\mu} - \epsilon_{\mu\alpha\beta}\oomega^{[\alpha\beta]} \quad . \label{Lambda} \eea

\no From (\ref{6.8})-(\ref{6.10}) and the gauge transformations (\ref{deltay3})-(\ref{deltase}) we have 

\bea \delta e_{\mu(\alpha\beta)} &=& \eta_{\alpha\beta} \Lambda_{\mu} - (\eta_{\alpha\mu}\Lambda_{\beta} +\eta_{\beta\mu}\Lambda_{\alpha})/2 + \frac 1{2\,m} \left\lbrack \epsilon_{\mu\alpha}^{\quad\nu}(\p_{\nu}\Lambda_{\beta}+\p_{\beta}\Lambda_{\nu}) + (\alpha \leftrightarrow \beta )\right\rbrack \nn \\
&+& \frac {\p_{\mu}}{2\, m^2} \left\lbrack \frac 23 \eta_{\alpha\beta} \p \cdot \Lambda - \p_{\alpha}\Lambda_{\beta} - \p_{\beta}\Lambda_{\alpha}\right\rbrack \quad . \label{6.15} \eea

\no which imply

\bea \delta (\eta^{\alpha\beta} e_{\mu(\alpha\beta)}) &=& \delta \, e_{\mu} = 2\, \Lambda_{\mu} \quad , \label{deltale} \\ \delta (\eta^{\alpha\beta} e_{\alpha(\beta\mu)}) &=& \delta \, \ooe_{\mu} = -\Lambda_{\mu} - \frac 1{m^2}\left( \Box\, \Lambda_{\mu} + \frac 13 \p_{\mu} \p \cdot \Lambda \right) \quad  \label{deltaleb} \eea

\no After extensive use of $\epsilon$-identities we have checked that (\ref{6.6}) is invariant under the Weyl transformations ({\ref{deltaA}) and (\ref{6.15}). For the opposite helicity we replace $(e_{\mu(\alpha\beta)},A_{\mu},\phi,\Lambda_{\mu}) \to (e_{\mu(\alpha\beta)}^-,A_{\mu}^-,\phi^-,\Lambda_{\mu}^-)$. This shows that the original doublet theory (\ref{lmp4}) is invariant under the gauge transformations associated with the 6 gauge parameters\footnote{Recall that $\omega_{\mu} = -2\, \epsilon_{\mu\alpha\beta}\oOmega^{\alpha\beta}$.} $(\oomega_{[\mu\nu]},\oOmega_{[\mu\nu]})$ corresponding to linear combinations of $(\Lambda_{\mu},\Lambda_{\mu}^-)$. The gauge symmetries associated with the remaining 19 independent gauge parameters $(\epsilon_{\mu},\epsilon_{(\alpha,\beta)},\oomega_{(\alpha\beta)},\oOmega_{(\alpha\beta)})$ are automatically implemented through the gauge invariant composite fields (\ref{6.8}), (\ref{6.9}) and (\ref{6.10}). As expected, all symmetries (\ref{deltay4})-(\ref{deltase})  hold true in the massive doublet theory (\ref{lmp4}). 

It is remarkable that the symmetries associated with $(\oomega_{[\mu\nu]},\oOmega_{[\mu\nu]})$ have become a non trivial dynamical symmetry involving second order time derivatives. Notice, in particular, that (\ref{6.15}) depends upon the symmetric combination $\p_{(\mu}\Lambda_{\nu )}$  while in $\delta A_{\mu}$ only  $\p_{[\mu}\Lambda_{\nu ]}$ appears, so there is no way of combining $e_{\mu(\alpha\beta)}$ with $A_{\mu}$  and its derivatives into a Weyl invariant tensor.

Regarding the particle content of (\ref{6.6}), how can we be sure that it correctly describes helicity $\pm 3 $ states ? This can be demonstrated by deriving the ghost free spin-3 self-dual model of \cite{ak1} from (\ref{6.6}), or equivalently, two opposite helicity copies of \cite{ak1} from the doublet model (\ref{lmp4}) via gauge fixing at action level. Indeed, the six gauges 

\be e_{\mu}^{\pm} =0 \quad , \label{gauge1} \ee

\no uniquely determine, see (\ref{Lambda}), (\ref{deltale}) and  footnote 3, the six parameters $(\oomega_{[\mu\nu]},\oOmega_{[\mu\nu]})$ thus, satisfying the ``completeness'' criteria of \cite{moto}. Therefore (\ref{gauge1}) can be fixed at action level without loosing relevant field equations. After breaking those symmetries,  from (\ref{deltaphiu}) we see that $\phi_{\mu}=0$ uniquely fix $\epsilon_{\mu}$. Analogously, $S_{(\alpha\beta)}=0=\obu_{(\alpha\beta)}=\oK_{(\alpha\beta)}$ uniquely fix $\epsilon_{(\alpha\beta)}, \oomega_{(\alpha\beta)}$ and $\oK_{(\alpha\beta)}$.  All those gauges amount to replace the composite field $e_{\mu(\alpha\beta)}$ by an elementary traceless field  
$\ooe_{\mu(\alpha\beta)} \,\, (\eta^{\alpha\beta}\ooe_{\mu(\alpha\beta)}=0) $ in (\ref{6.6}) so reproducing the model \cite{ak1}.

 Alternatively, instead of (\ref{gauge1}),  we might have fixed  $a_{\mu}^{\pm} \equiv A_{\mu}^{\pm} \mp \epsilon_{\mu\alpha\beta}\p^{\alpha}e^{\beta}_{\pm}/(4\, m)=0$ and get rid of the auxiliary fields $A_{\mu}$ and $\phi$ (which decouples) while keeping the traces $e_{\mu}^{\pm}\ne 0$.   Notice that $\delta a_{\mu}^{\pm} = 3\, \Lambda_{\mu}^{\pm}/2$ which guarantees that $a_{\mu}^{\pm}=0$ uniquely fix the three parameters $\Lambda_{\mu}^{\pm}$, so it can be fixed at action level without problems. The price we pay for eliminating $A_{\mu}^{\pm}$ is the presence of third order terms in derivatives of $e_{\mu}^{\pm}$.
 
\section{Conclusion}

Here we have shown how to  obtain first order spin-s self dual models (parity singlets) in $D=2+1$ via Kaluza-Klein dimensional reduction of the Vasiliev's \cite{vasiliev} first order action for massless spin-s particles in  $D=3+1$. We have explicitly worked out the cases $s=1,2,3$. For $s=1,2$ we reproduce the self dual models of \cite{tpn} and \cite{ak} respectively, while at $s=3$ we obtain a new self dual model (\ref{6.6}) invariant under the non trivial local Weyl transformations (\ref{deltaA}) and (\ref{6.15}). After fixing the gauge $e_{\mu}=0$ we recover the model of \cite{ak1}.

 There are several questions about the Weyl symmetry that we are currently addressing. Namely,  its importance for the definition  of the self dual model itself and for the introduction of cubic and higher order self interacting vertices for  spin-3 particles in $D=2+1$. 
 
 Even the origin of the symmetry is not yet clear.  Usually, the KK dimensional reduction leads to Stueckelberg fields that can eliminated via an unitary gauge such that no local symmetry is left over, since the number of independent gauge parameters equals the number of new fields. For instance, in the spin-1 case $e_D$ is gauged away by the $U(1)$ symmetry while in the spin-2 case the 10 fields $(e_{\mu D},e_{D\mu},e_{DD},y_{[\mu\nu]D})$ can be eliminated
 by the 10 gauge parameters $(\epsilon_{\mu},\oomega_{\mu},\epsilon,\omega_{\mu})$. However, in the spin-3 case the 25 independent\footnote{Recall that $\oomega_{\mu(\nu D)}=\oomega_{\mu\nu}$ is traceless, see (\ref{tr3}), and the 10 constraints (\ref{omegac}) reduce the 18 variables $\omega_{\mu(\alpha\beta)}$ to 8 independent parameters.} parameters $(\xi_{(\mu D)},\xi_{(\alpha\beta)},\oomega_{\mu(\nu D)},\omega_{\mu(\alpha\beta)})$ are not enough to eliminate the 33 new fields $(e_{D(D\mu)},e_{D(\alpha\beta)},e_{\mu(\nu D)}, y_{[\mu\nu](\alpha\, D)},e_{\mu}, \epsilon_{\mu}^{\,\,\,\rho\gamma}y_{[\rho\gamma](\alpha\alpha)})$, there are 8 exceeding fields. They correspond to the necessary auxiliary fields $(\phi^{\pm},A_{\mu}^{\pm})$ which are linear combinations of the antisymmetric components of the tensors $e_{\mu(\nu D)}$ and $ y_{[\mu\nu](\alpha\, D)}$. Therefore, all remaining new fields must be eliminated, including the vectors $(e_{\mu}, \epsilon_{\mu}^{\,\,\,\rho\gamma}y_{[\rho\gamma](\alpha\alpha)})$, or equivalently $e_{\mu}^{\pm}$. However, we have not been able to foresee that the Weyl symmetry associated with the elimination of $e_{\mu}^{\pm}$ would be dynamically realized. In the $s=1$ and $s=2$ cases we are able to define  composite fields invariant under all gauge transformations. This is apparently not possible in the spin-3 case.
 A similar situation occurs in the KK dimensional reduction of  massless limit of NMG from $D=2+1$ to $D=1+1$, see \cite{bdg}.
 
 We stress that differently from the $s=1$ and $s=2$ cases, for $s=3$ it is not clear which new fields are pure gauge and could be eliminated right from the start in order to avoid the several field redefinitions  that we have done. In particular, some of the new fields correspond to $(A_{\mu}^{\pm}, \phi^{\pm})$ and must be kept in the off-shell formulation of the theory.

Hopefully the analysis of the next case
 ($s = 4$) will clarify the issue of dynamical versus trivial Stueckelberg symmetries. In the general integer spin-s case the action of \cite{vasiliev} in $D=3+1$ depends on two fields $S[\oy_{[MN](A_1 \cdots A_{s-1})},\ooe_{M(A_1 \cdots A_{s-1})}]$. After the dimensional reduction, the field redefinitions (\ref{f1}),(\ref{f2}) and (\ref{f3}) will be replaced by $y_{[\mu,\nu](\alpha_1 \cdots \alpha_{s-1})}=m\, \epsilon_{\mu\nu}^{\quad\,\rho}f_{\rho(\alpha_1 \cdots \alpha_{s-1})} $. So  the helicity eigenstates will be linear combinations of $f_{\rho(\alpha_1 \cdots \alpha_{s-1})}$ and $e_{\rho(\alpha_1 \cdots \alpha_{s-1})} $. This is now in progress.
 We point out that $s=3$ is still special. The next case $s=4$ is more promising regarding an arbitrary integer spin generalization.
 
Finally, we hope that the new model (\ref{6.6}) will allow us to run the successive Noether gauge embeddings in order to derive the whole sequence of self dual models $SD_j^{(3)}$ ($j=1,2, \cdots,6$) without getting stuck at $SD_4^{(3)}$. The issue of soldering of opposite helicities and the connection with $D=3+1$ massive models as well as the generalization of (\ref{6.6}) and the Weyl symmetry to curved space backgrounds is also under investigation.

\section{Acknowledgements}

This work is partially supported by CNPq  (grant 306380/2017-0).

\end{document}